\documentclass[12pt]{article}
\usepackage{graphicx}

\usepackage{cite}

\begin{document}

\title{Opinion Dynamics and Unifying Principles: \\ A Global Unifying Frame}

\author{ Serge Galam\thanks{serge.galam@sciencespo.fr} \\
CEVIPOF - Centre for Political Research, Sciences Po and CNRS,\\
1, Place Saint Thomas d'Aquin, Paris 75007, France}

\date{January 29, 2020}
\maketitle

\begin{abstract}

I review and extend the set of unifying principles, which allow comparing all models of opinion dynamics within one single frame. Within the Global Unifying Frame (GUF), any specific update rule chosen to study opinion dynamics for discrete individual choices is recast into a probabilistic update formula. The associated dynamics is deployed using a general probabilistic sequential process, which is iterated via the repeated reshuffling of agents between successive rounds of local updates. The related driving  attractors and tipping points are obtained with non-conservative regimes featuring both threshold and threshold-less dynamics. Most stationary states are symmetry broken, but fifty-fifty coexistence may also occur. A practical procedure is exhibited for several versions of Galam and Sznajd models when restricted to the use of three agents for the local updates. Comparing these various models, some are found to be identical within the GUF. Possible discrepancies with numerical simulations are discussed together with the difference between the GUF procedure and a mean field approach.

\end{abstract}

Key words: Sociophysics, opinion dynamics, local updates, unifying rules

\section{Opinion dynamics, models and reality}

Opinion dynamics is a major topic of sociophysics with a series of different models being used \cite{brazil, frank, book, fortu, bikas, rum}. Each one of these models accounts for a peculiar social feature, which in turn drives social exchanges among a group of people. The chosen mechanism gives rise to a specific local update rule to monitor the related changes between the competing opinions. Accordingly, the relevance of a given model is rooted in the social principle underlying the dynamics. The related social principle legitimates the model and therefore, the associated results \cite{stauffer-sn}.

Often, numerical simulations are required to study the dynamics produced by iterating the update rule. Analytical calculations are scarce. In addition, validation with real data is mostly not directly applicable. The goal is to identify trends to make some predictions on real events to come like elections. 

On this basis, the focus of each model lies on the social mechanism enlightened to set the update local rule. The associated results obtained are then argued quite naturally as being the consequence of the chosen social mechanism. And in turn those mechanisms are put forward to explain the social outcome.

Although the above approach is completely legitimate, it carries an underlying limitation due to the lack of certainty about the outcomes obtained using simulations with possible long relaxation times and trapping into local minima. Therefore, associated conclusions about the targeted social reality may turn misleading. In particular, wrong social or political strategies could be selected to intervene on reality. 

Accordingly, to avoid such possible social mistakes I was able to identify unifying principles, which allow building a Global Unifying Frame (GUF) to review and compare all models of opinion dynamics using discrete individual choices \cite{uni}. Within the GUF any specific update rule can be recast into a probabilistic update formula.

A general probabilistic sequential process is then implemented through a series of successive local updates separated in between by repeated reshuffling of agents. The probabilistic update formula yields the attractors and tipping points that drive the related dynamics. Non conservative regimes featuring both threshold and threshold-less dynamics are obtained with mostly symmetry broken stationary states but fifty-fifty coexistence may also occur. 

A practical procedure to implement the GUF is exhibited for several versions of Galam and Sznajd models when the local updates are restricted to using three agents. Some of the respective probabilistic update formulas are found to be identical pointing that the social meaning of different principles of opinion updates are indeed irrelevant to the outcomes. This founding puts at stake the claim that a given result would be directly linked to the social feature put forward to legitimate the update rule.

Possible discrepancies between the GUF outcomes with numerical simulations from the models are discussed together with the difference between the GUF procedure and a mean field approach.

This work subscribes to a series of papers dealing with the same goal of building tools to allow comparing the various models of opinion dynamics \cite{c1, c2, c3}.

The rest of the paper is organized as follows: Section introduces the issue of updating rules validated by a claimed social mechanism. The general probabilistic frame is developed in Section 2, while Section 3 accounts for iterating the dynamics to get the Global Unifying Frame (GUF). Section 4 applies the GUF to a series of Galam and Sznajd opinion models. A discrepancy between the GUF and other works is solved in Section 5. Section 6 discusses mean-filed versus the GUF, and the last Section contains concluding comments.

\section{The general probabilistic frame}

I consider a population of $N$ agents $\{S_{l,t}\}$ at time $t$ with $l=1, 2, ... , N$. I assume that each agent holds an individual discrete choice $\pm$ with $S_{l,t}=+$ or $S_{l,t}= -$. The respective proportions of agents sharing opinion $+$ and $-$ at time $t$ are denoted $p_{t}$ and $(1-p_{t})$.

The opinion dynamics is driven by first distributing randomly all agents $\{S_{l,t}\}$ at time $t$  in groups of size $r$. Then an update rule is applied  simultaneously to each group yielding at time $t+1$ a new set $\{S_{l,t+1}\}$. Without loss of generality $N$ is chosen to be a multiple of $r$. Focusing on the case $r=3$ yields $2^3=8$ possible configurations for each group of three agents with,

\begin{itemize} 
\item $a_{1}= + + +$ with probability $p_{t,1}$ ,
\item $a_{2}= + + - $ with probability $p_{t,2}$ ,
\item $a_{3}= + - +$ with probability $p_{t,3}$ ,
\item $a_{4}= - + + $ with probability $p_{t,4}$ ,
\item $a_{5}= - - + $ with probability $p_{t,5}$ ,
\item $a_{6}= - + -$ with probability $p_{t,6}$ ,
\item $a_{7}= + - -$ with probability $p_{t,7}$ ,
\item $a_{8}= - - - $ with probability $p_{t,8}$ .
\end{itemize}

When all configurations of agents are contributing to the update, the probabilities for the eight configurations $a_i$ are respectively:

\begin{itemize} 
\item $p_{t,1}= p_{t}^3$
\item $p_{t,2}=p_{t,3}=p_{t,4}=p_{t}^2 (1-p_{t})$
 \item $p_{t,5}=p_{t,6}=p_{t,7}=p_{t}(1-p_{t})^2$
 \item $p_{t,8}=(1-p_{t})^3$.
\end{itemize}

It is of importance to stress that the procedure can also account  for peculiar distributions of agents. For instance, correlations in the initial distribution of agents can be included like with antiferromagnetic like arrangement $+ - + -+ -+ - +....$. In this case some configurations are excluded from the update which imposes to set $p_{t,1}= p_{t,2}=p_{t,4}=  p_{t,5}=p_{t,7}=p_{t,8}= 0$ and $p_{t,3}= p_{t,6}=1/2$. Other specific configurations can also be discarded implying then a renormalization of the probabilities of the contributing ones to get their sum equal to one. Another illustration is given in Section \ref{dis}.

Depending on the update rule and the local configurations, each agent at time $t+1$ either preserves its opinion from time $t$ or shifts to the opposite opinion. Each model of opinion dynamics sets a specific update rule,  which associates a configuration $b_i$ to each configuration $a_i$ with the series $a_{i} \rightarrow b_{i}$ for $i=1, 2, ..., 2^r$ where $a_{i}=S_{1,t}, S_{2,t}, ..., S_{r,t}$ and $b_{i}=S_{1,t+1}, S_{2,t+1}, ..., S_{r,t+1}$.

The content of the update rule must specify which agents among the $r$ are being updated and how when applying the local rule. For instance, one can consider that for each configuration  $b_i$ either the $r$ agents have been updated or only the one on the left side, or on the right side,  or a pair or any other choice. This indication matters for both calculating $p_{t+1}$ analytically and performing Monte Carlo steps in simulations.

What counts is the number of agents being updated in the model, not the number of agents shifting opinion. For instance, in the seminal Galam model (Subsection \ref{lmm}) the three agents are updated, which does not prevent some agents to keep their opinion as for the two $+$ in the configuration $+-+$ that transforms to $+++$. In contrast, in the original Sznajd model (Subsection \ref{oom}), only one single agent is updated as seen with both configurations $++(+)$ and $++(-)$, which yield respectively $++(+)$ and $++(+)$. There, only the last agent on the right side is updated.

For groups of size $r$ with a given update rule, the number of agents $r_u$ being updated in the various configurations may vary from $1$ to $r$ depending on the model specification. Then,  for each updated configuration $b_i$, I denote $k_i$ the number of $+$ among the $r_u$ agents with $k_i=0, 1, ..., r_u$. Therefore, each configuration $b_i$ has a proportion $k_i/r_u$ of $+$ and $(r_u-k_i)/r_u$ of $-$. 

In addition, the update rule $a_i\rightarrow b_i$ may also be probabilistic with $V$ possible outcomes $b_{i,v}$ associated with respective probabilities $\alpha_v$ satisfying $\sum_{v=1}^V\alpha_v=1$. In that case, the number $k_i$
of $+$ has to be replaced by the average number of $+$ over the V outcomes with,
\begin{equation}
\bar{k}_i=\sum_{v=1}^V\alpha_v k_{i,v} ,
\label{k}
\end{equation}
giving an averaged proportion $\bar{k}_i/r$ of $+$ for configuration $i$. 

Then, I calculate the probability $p_{t+1}$ that an agent selected randomly holds opinion $+$ at time $(t+1)$ from an initial  $p_{t}$  at time $t$ by adding all contributions to the $+$ opinion from each configuration $b_i$, which yields the general expression,
\begin{equation}
p_{t+1}= \frac{1}{r_u} \sum_{i=1}^{2^r} \bar{k}_{i} p_{t,i} ,
\label{u1}
\end{equation}
where $p_{t,i}$ is the probability to have configuration $i$ from the $2^r$ possible configurations of $+$ and $-$ for a group of $r$ agents with $r_u$ being updated.

For $r=3$ with no exclusion of configurations, Eq. (\ref{u1}) writes,
\begin{eqnarray}
\nonumber p_{t+1}&=&\frac{1}{r_u} \big\{k_{1} p_{t}^3 \\
\nonumber& + & 
(k_{2} + k_{3} + k_{4}) p_{t}^2 (1-p_{t}) \\
\nonumber& + & 
(k_{5} +k_{6} +k_{7}) p_{t}(1-p_{t})^2 \\
& + & 
k_{8} (1-p_{t})^3  \big\},
\label{u3}
\end{eqnarray}
with $k_i\rightarrow \bar{k}_{i}$ for a probabilistic case.

\section{Iterating the dynamics: the Global Unifying Frame}

Eq. (\ref{u1}) yields the new proportion of $+$ at time $t+1$ knowing the proportion $p_t$ at time $t$. This change of the proportion of $+$ results from one update of opinions obtained applying a local update rule to all groups of $r$ agents selected randomly. To launch a dynamics I iterate the process by first, breaking down the groups, second, reshuffling all agents, third, redistributing randomly agents in groups of size $r$, fifth, applying again the local update rule. That generates a new proportion $p_{t+2}$ from $p_{t+1}$. After $n$ successive updates I get the series $p_{t}\rightarrow p_{t+1}\rightarrow p_{t+2}\rightarrow ... \rightarrow p_{t+n}$.

The instrumental question is then to find out if the dynamics converges towards some attractor after a given number $n_a$ of updates with $p_{t+n_a-1} \rightarrow p_{t+n_a} \approx p_{t+n_a+1}$. The answer is obtained solving the fixed point Equation,
\begin{equation}
p_t= \frac{1}{r_u} \sum_{j=1}^{2^r} k_{i} p_{t,j} ,
\label{f1}
\end{equation}
which yields all attractors and thresholds driving the dynamics monitored but the used update local rule. For a probabilistic case $k_i\rightarrow \bar{k}_{i}$.  Eqs. (\ref{u1}, \ref{f1}) defines the Global Unifying Frame (GUF).

For $r\geq 5$ a numerical solving is required while the cases $r=2, 3, 4$ allow some analytical solving. In the case $r=3$ with all configurations contributing,  Eq. (\ref{f1}) writes,

\begin{equation}
p_{t}=\frac{1}{r_u}  \Big \{ a p_t^3+ b p_t^2+ c p_t+d \Big \}  ,
\label{u5}
\end{equation}
with 
\begin{eqnarray}
\nonumber a&=&  k_ 1-k_ 2-k_ 3-k_ 4+k_ 5+k_ 6+k_ 7-k_ 8 \\
\nonumber b& = & k_ 2+k_ 3+k_ 4-2 k_ 5-2 k_ 6-2 k_ 7+3 k_ 8  \\
\nonumber c& = & k_ 5+k_ 6+k_ 7- 3 k_ 8 \\
d& = & k_8 .
\label{f3}
\end{eqnarray}
It is a cubic Equation with,
\begin{equation}
a p_t^3+ b p_t^2+ (c -r_u) p_t+d=0 ,
\label{f2}
\end{equation}
and the associated discriminant,
\begin{equation} 
D =  18 a b (c -r_u) d -4 b^3 d + b^2 (c -r_u)^2 -4 a (c -r_u)^3 -27 a^2 d^2 ,
\label{f8}
\end{equation}
determines the number of real solutions. Three cases are possible, 

\begin{enumerate}
\item $D>0$: Eq. (\ref{f2}) has are three real roots $p_{c,1}<p_{c,2}<p_{c,3}$ which are distinct. Dealing with proportions these roots are  acceptable only when satisfying the condition $0\leq p_{c,i} \leq 1$ for $i=1, 2, 3$.  By symmetry that happens  either for the three of them with one separator and two attractors or for only one of them which is then an attractor. 

For the first scenario, when $p_t<p_{c,2}$, the update iteration drives the opinion $+$ towards $p_{c,1}$ leading to the victory of opinion $-$. At opposite, for $p_t>p_{c,2}$, the update iteration drives the opinion $+$ towards $p_{c,3}$ leading to its victory. One case is shown in the upper part of Figure (\ref{fig1}) with $p_{c,2}<1/2$. The opposite $p_{c,2}>1/2$ may also occur. Each attractor can be a pure phase or a mixed phase with respectively $p_{c,1}=0$ or $p_{c,1}>0$ and $p_{c,3}=1$ or $p_{c,1}<1$.

\begin{figure}
\centering
\includegraphics[width=1\textwidth]{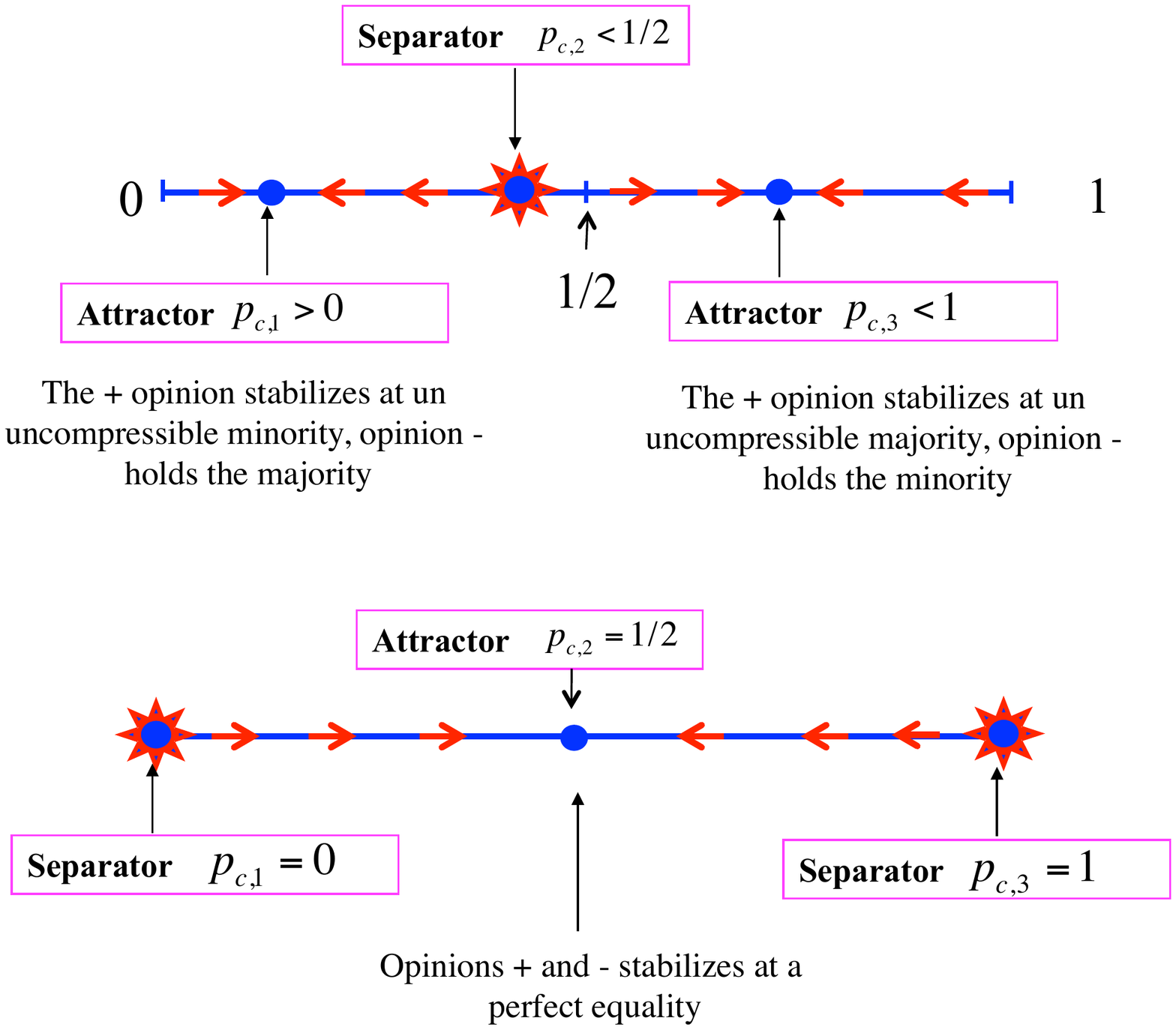}
\caption{Case $D>0$ with three fixed points. Upper part of the Figure shows a threshold dynamics with two attractors $p_{c,1}$ and $p_{c,3}$ separated by an unstable fixed point $p_{c,2}$. A case with $p_{c,2}\leq 1/2$, $0\leq p_{c,1}< p_{c,2}$ and $p_{c,2}<p_{c,3}\leq 1$ is exhibited. Lower part of the Figure shows a threshold dynamics with two separators $p_{c,1}$ and $p_{c,3}$ with an attractor $p_{c,2}$ in between. In such a case $p_{c,1}=0$, $0<p_{c,2}< 1$ and $p_{c,3}=1$.
}
\label{fig1}
\end{figure}    

In case $p_{c,1}=0$ and $p_{c,3}=1$, the fixed point $p_{c,2}$ can also become an attractor with both pure ones being unstable in its direction as shown in the lower part of Figure (\ref{fig1}). In that case, due to topological constraint, the transition from $p_{c,2}$ being a separator to $p_{c,2}$ being an attractor occur via a conservative regime $p_{t+1}=p_t$ where each point is a fixed point, thus recovering the voter model as shown in Figure (\ref{fig2}). It is worth noting that in case $p_{c,1}>0$ and $p_{c,3}<1$ such a transformation of a separator into an attractor is still possible but now there is no transition via the voter model. An illustration of both cases is given in Section (\ref{ap}).

\begin{figure}
\centering
\includegraphics[width=1\textwidth]{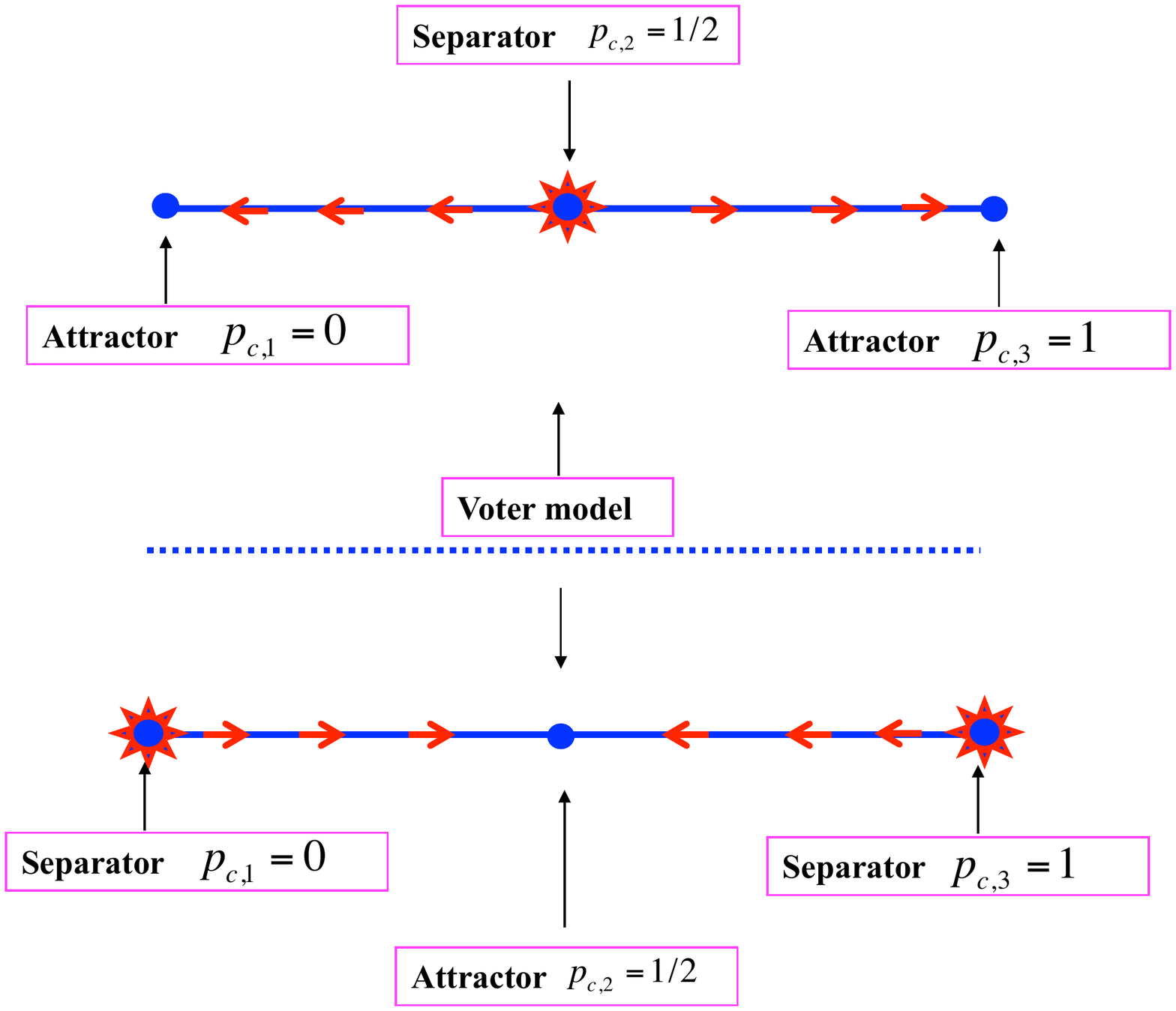}
\caption{The transformation of a threshold dynamics with the two attractors $p_{c,1}=0$ and $p_{c,3}=1$ and the separator $p_{c,2}=1/2$ (higher part) into a threshold-less like dynamics with the two separators $p_{c,1}=0$ and $p_{c,3}=1$ and the attractor $p_{c,2}=1/2$ (lower part). It must pass via a voter model (middle part) where each point is conserved by the dynamics.}
\label{fig2}
\end{figure}  

\item $D=0$: Eq. (\ref{f2}) exhibits three real roots with at least a double one which corresponds to a transition regime from threshold dynamics to threshold-less dynamics. Two subclasses can occur: 

\begin{description}
\item[(i)] One single attractor and a double fixed point which is an attractor on one side and separator on the other side, the side where the attractor is located. The upper part of Figure (\ref{fig3}) exhibits the case $p_{c,1}=p_{c,2}$ with $p_{c,3}$ being the attractor. The symmetric case is also possible with $p_{c,1}$ being the attractor and $p_{c,2}=p_{c,3}$.

\item[(ii)] Or a triple attractor $p_{c,1}=p_{c,2}=p_{c,3}$ making the dynamics threshold-less. Whatever the initial conditions are, the repeated updates drive the collective opinion towards the single attractor. Two cases are possible. The first one has $p_{c,1}=p_{c,2}=p_{c,3}=1/2$, which means the dynamics leads to a coexistence phase with a perfect fifty/fifty equality. The second one is not balanced with $p_{c,1}=p_{c,2}=p_{c,3}\neq 1/2$, which means the dynamics leads to a a stable majority / minority coexistence phase with a deterministic victory for one specific opinion. if $p_{c,1}=p_{c,2}=p_{c,3}< 1/2$ opinion $+$ is certain to lose provided some number of updates are completed. Otherwise when $p_{c,1}=p_{c,2}=p_{c,3}> 1/2$ opinion $+$ wins the competition. One case with $p_{c,1}=p_{c,2}=p_{c,3}> 1/2$ is shown in the lower part of Figure (\ref{fig3}). The symmetric situation with $p_{c,1}=p_{c,2}=p_{c,3}< 1/2$ is also possible making opinion $+$ to lose the competition. 
\end{description}

\begin{figure}
\centering
\includegraphics[width=1\textwidth]{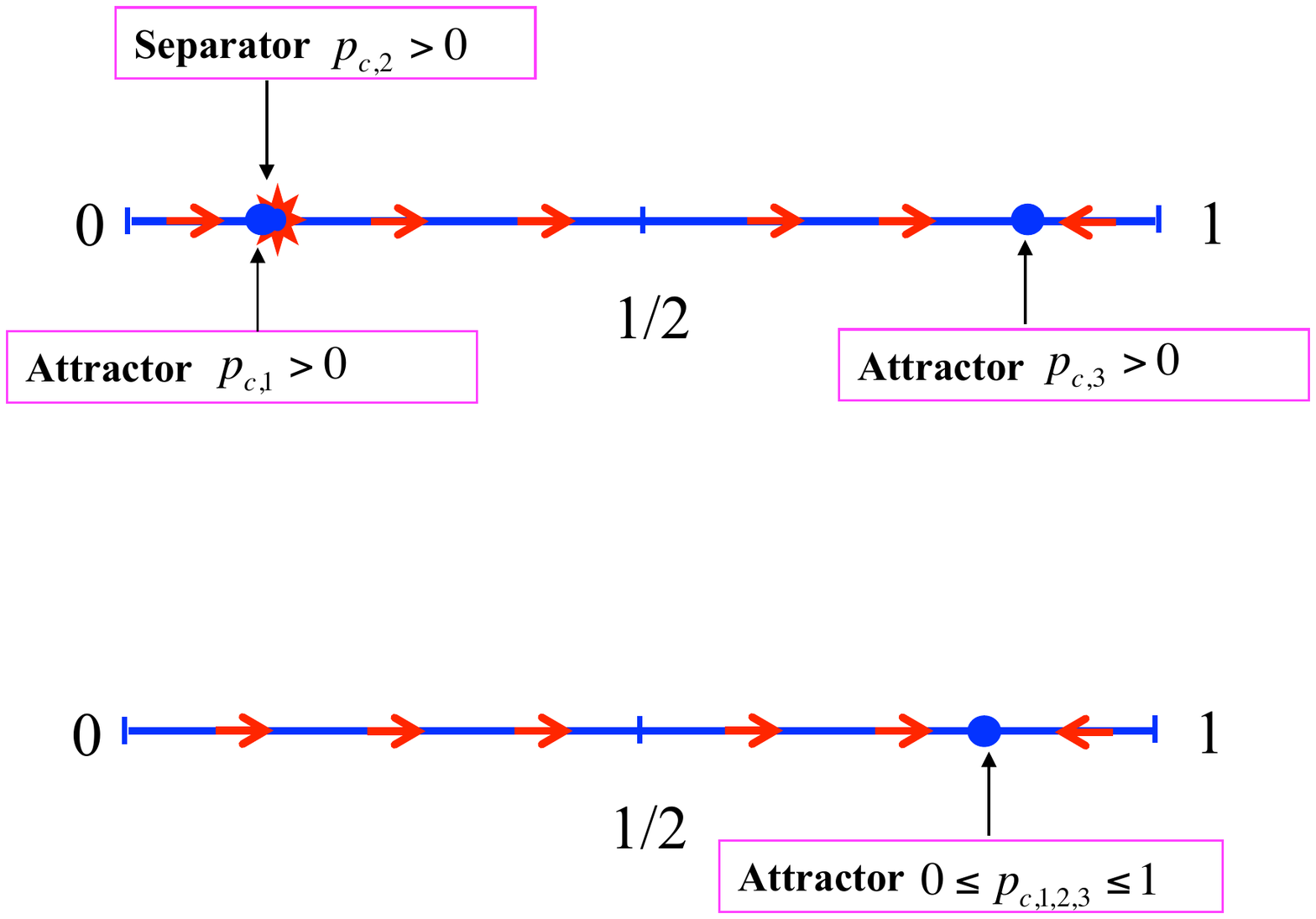}
\caption{Upper part of the Figure shows a case $D=0$ with a double fixed points $p_{c,1}=p_{c,2}$, which is attractor below it and separator above it. The attractor of the dynamics is $p_{c,3}$. The Case $D=0$ with a single triple fixed point $p_{c,1}=p_{c,2}=p_{c,3}$ which is an attractor is shown in lower part of the Figure. What ever the initial $p_t$ is the dynamics leads towards $p_{c,1}=p_{c,2}=p_{c,3}$ to reach it provided the required number of updates has been performed. Otherwise it stops before. The attractor can be located anyplace with $0\leq p_{c,1}=p_{c,2}=p_{c,3} \leq 1$.
}
\label{fig3}
\end{figure}    

\item $D<0$: Eq. (\ref{f2}) has one real root and two imaginary roots. It is thus a one single attractor dynamics. The attractor can be located at any value between $0$ and $1$ depending on the details of the update rule.
\end{enumerate}

\section{Applying the GUF to existing opinion dynamics models}
\label{ap}

The Global Unifying Frame can in principle be applied to any update rule used for a two state opinion dynamics model. To demonstrate my claim, in the following, I apply the GUF to a series of Galam and Sznajd models restricted to $r=3$. I first compare what the scheme yields with respect to the results obtained directly from the respective models. Then the scheme allows a comparison between the two series of models. For each model, I need to evaluate $b_i$ and $\bar{k_i}$ in order to calculate the update equation and the related discriminant $D$ from Eq. (\ref{f8}) to determine the associated dynamics. 

\subsection{Application to Galam models}

\subsubsection{The Local Majority Model (LMM) \cite{voting, chopard, mino}}
\label{lmm}

The basis set of Galam model considers a population of agents, who are randomly randomly distributed in groups of size $r$. A local majority rule is then applied simultaneously to each group. In case of a tie at an even size group a tie breaking rule is applied.  Afterwards all agents are reshuffled and the previous scheme is iterated again. And so on and forth. 

When all groups are of size $r=3$, $r_u=3$. All $8$ configurations are thus contributing to the update with $p_{t,1}= p_{t}^3$, $p_{t,2}=p_{t,3}=p_{t,4}=p_{t}^2 (1-p_{t})$, $p_{t,5}=p_{t,6}=p_{t,7}=p_{t}(1-p_{t})^2$, $p_{t,8}=(1-p_{t})^3$. It yields respectively, 
\begin{itemize} 
\item $b_1=b_2=b_3=b_4=+++$,
\item $b_5=b_6=b_7=b_8=---$,
\end{itemize}
with,
\begin{itemize}
\item $k_1=k_2=k_3=k_4=3$,
\item $k_5=k_6=k_7=k_8=0$,
\end{itemize}
and $a=-6, b=9, c=0, d= 0$, making Eqs. (\ref{u5}) and (\ref{f8}) to write, 
\begin{equation}
p_{t}=-2p_t^3+3p_t^2 ,
\label{a1}
\end{equation}
and
\begin{equation}
D=81>0 ,
\label{aa1}
\end{equation}
which implies three real roots $p_{c,1}<p_{c,2}<p_{c,3}$, which are $p_{c,1}=0, p_{c,2}=1/2, p_{c,3}=1$. 

Their respective stabilities are determined by the sign of $\lambda_{c,i} -1$ with,
\begin{equation}
\lambda_{c,i}=\frac{\partial p_{t+1}}{\partial p_{t}} \bigg |_{p_{c,i}} ,
\label{l1}
\end{equation}
which gives using Eq. (\ref{a1}),
\begin{equation}
\lambda_{c,i}=6p_{c,i}(1-p_{c,i}) ,
\label{l2}
\end{equation}
with $i=1, 2, 3$. 

Instead of the sign of $\lambda_{c,i} -1$ it is more convenient to check if $\lambda_{c,i}>1$ or $\lambda_{c,i}<1$.  In the first case, the fixed point is a separator while in the second case it is an attractor. Having $\lambda_{c,1}=\lambda_{c,3}=0$ and $\lambda_{c,2}=3/2$ makes the three fixed points $p_{c,1}, p_{c,2}, p_{c,3}$  respectively attractor, separator, attractor.

At this point, I underline the fact that solving the LMM directly is quasi identical to applying the GUF since indeed the GUF is a generic extension of what I have been implementing in my models of opinion dynamics making here no surprise with the findings.

\subsubsection{ The Contrarian Majority Model (CMM) \cite{contra}}

The Contrarian Majority Model proceeds first precisely as the LMM. However, after applying the local majority rules to each group of size $r$r, a fraction $\alpha$ of agents shifts their individual opinion to the opposite opinion one. Accordingly, only the proportion $(1-\alpha)$ of agents keeps the opinion they got from the local majority choice and a proportion $\alpha$ holds the opposite ones. Then, all agents are reshuffled and the two-step previous scheme is repeated. And so on and forth. 

In the case with $r=3$, $r_u=3$ as for the LMM with all $8$ configurations contributing to the update. However, here I have to consider the two cases:

\begin{itemize}
\item Majority rule yielding,
\begin{itemize}
\item $b^1_1=b^1_2=b^1_3=b^1_4=+++$ ,
\item $b^1_5=b^1_6=b^1_7=b^1_8=---$ ,
\end{itemize}
and
\begin{itemize}
\item $k^1_1=k^1_2=k^1_3=k^1_4=3$ ,
\item $k^1_5=k^1_6=k^1_7=k^1_8=0$ ,
\end{itemize}
with probability $(1-\alpha)$.

\item Contrarian shifts yielding,
\begin{itemize}
\item $b^2_1=b^2_2=b^2_3=b^2_4=---$ ,
\item $b^2_5=b^2_6=b^2_7=b^2_8=+++$  ,
\end{itemize}
and
\begin{itemize}
\item $k^2_1=k^2_2=k^2_3=k^2_4=0$ ,
\item $k^2_5=k^2_6=k^2_7=k^2_8=3$ .
\end{itemize}
with probability $\alpha$.

Then the two cases must be averages giving,
\item Average 
\begin{itemize} 
\item $\bar{k}_1=\bar{k}_2=\bar{k}_3=\bar{k}_4=3 (1-\alpha)$ ,
\item $\bar{k}_5=\bar{k}_6=\bar{k}_7=\bar{k}_8=3 \alpha$ , 
\end{itemize}
\end{itemize}
which yields to $a=-6(1-2\alpha), b=9(1-2\alpha), c=0, d= 3\alpha$ for Eq. (\ref{u5}) with,
\begin{equation}
p_{t}=-2(1-2\alpha) p_t^3+3(1-2\alpha) p_t^2 +\alpha ,
\label{a2}
\end{equation}
and,
\begin{equation}
D=81 (-1 + 2 \alpha) (-1 + 6 \alpha)^3 ,
\end{equation}
for Eq. (\ref{f8}), which can be either negative, null or positive depending on $\alpha$. The three cases are:

\begin{itemize} 
\item $\alpha < 1/6 \rightarrow D>0 \rightarrow$ three real roots $p_{c,1}<p_{c,2}<p_{c,3}$.
\item $1/6<\alpha < 1/2 \rightarrow D<0 \rightarrow$ a threshold-less dynamics with one single attractor.
\item $\alpha >1/2  \rightarrow D>0 \rightarrow$ again three real roots with $\alpha >1/2$ implying an oscillating regime.
\end{itemize}

Eq. (\ref{a2}) yields the three solutions,
\begin{eqnarray}
p_{c,1}&=&   \frac{-1 + 2 \alpha + \sqrt{1 - 8 \alpha + 12 \alpha^2}}{2 (-1 + 2 \alpha)}, \\
p_{c,2}& = & \frac{1}{2},  \\
p_{c,3}& = & \frac{-1 + 2 \alpha - \sqrt{1 - 8 \alpha + 12 \alpha^2}}{2 (-1 + 2 \alpha)} ,
\label{f3}
\end{eqnarray}
which requires  $\alpha \leq1/6$ or  $\alpha \geq1/2$ to exist. Otherwise $p_{c,2}=\frac{1}{2}$ is the unique solution.

Determining the stability of $p_{c,2}$ is sufficient to determine the respective dynamics of above three cases. From Eq. (\ref{l1}) using Eq. (\ref{a2}) I get,
\begin{equation}
\lambda_{c,i}=6 (1 - 2 \alpha ) (1 - p_{c,i}) p_{c,i} ,
\label{l3}
\end{equation}
which yields $3/2 - 3\alpha$ at  $p_{c,2}$. Therefore  $p_{c,2}$ is a separator when $3/2 - 3\alpha>1$, which is satisfied when $\alpha <1/6$. Otherwise, when $\alpha >1/6$, $p_{c,2}$ is an attractor since $3/2 - 3\alpha>1$.

When $\alpha <1/6$, $p_{c,2}$ being a separator,  both $p_{c,1}$ and $p_{c,3}$ are attractors. In contrast, for $\alpha >1/2$, $p_{c,2}$ is an attractor and thus both $p_{c,1}$ and $p_{c,3}$ are separators and as such must be located out of the ``physical range", i.e., outside the range $0,1$. Indeed $\alpha >1/2$ creates an oscillatory regime towards $p_{c,2}$. 

\subsubsection{The Extended Majority Model (EMM) \cite{red1}}

Mobillia and Redner extended the Local Majority Model in the case of groups of size three by allowing the possibility of one agent to influence two others \cite{red1}. While the configurations $+++$ and $---$ still yield $+++$ and $---$ as in the LMM, the configuration $++-$ becomes $+++$ with a probability $(1-\beta)$ and $---$ with a probability $\beta$. Similarly the configuration $--+$ becomes $---$ with probability $(1-\beta)$ and $+++$ with a probability $\beta$.

Therefore, as for the LMM, $r_u=3$  and all $8$ configurations contribute to the update rule. Similarly to the CMM, two cases must be included in the calculation:

\begin{itemize} 
\item Probability $(1-\beta)$ 

\begin{itemize} 
\item $b^1_2=b^1_3=b^1_4=+++$ ,
\item $b^1_5=b^1_6=b^1_7=---$ ,

\end{itemize}
yielding,
\begin{itemize}

\item $k^1_2=k^1_3=k^1_4=3$ ,
\item $k^1_5=k^1_6=k^1_7=0$. 

\end{itemize}

and with,

\item Probability $\beta$ 

\begin{itemize} 
\item $b^2_2=b^2_3=b^2_4=---$ ,
\item $b^2_5=b^2_6=b^2_7=+++$ ,
\end{itemize}
yielding,
\begin{itemize}
\item $k^2_2=k^2_3=k^2_4=0$ ,
\item $k^2_5=k^2_6=k^2_7=3$ ,
\end{itemize}

Both cases lead to the,

\item Averages

\begin{itemize} 
\item $\bar{k}^1_2=\bar{k}^1_3=\bar{k}^1_4=3 (1-\beta)$ ,
\item $\bar{k}^1_5=\bar{k}^1_6=\bar{k}^1_7=3 \beta$ .
\end{itemize}
\end{itemize}

Adding $b_1=+++, b_8=---$ with  $k_1=3, k_8=0$ completes the update of the eight configurations $a_i \rightarrow b_i$ of three agents, which allows to get $a=-6(1-3\beta), b=9(1-3\beta), c=9 \beta, d= 0$ for Eq. (\ref{u5}) with,
\begin{equation}
p_{t}=-2(1-3\beta) p_t^3+3(1-3\beta) p_t^2 +3\beta p_t ,
\label{a3}
\end{equation}
and,
\begin{equation}
D=81 (1-3 \beta)^4 ,
\end{equation}
 for for Eq. (\ref{f8}) which is always positive and null for $\beta =1/3$. 

For $\beta \neq1/3$, Eq. (\ref{a3}) yields the three fixed points $p_{c,1}=0, p_{c,2}=1/2, p_{c,3}=1$ as for the LMM case. However the main difference relates to the stability of $p_{c,2}=1/2$ with,
\begin{equation}
\lambda_{c,i}=6 (1 - 3 \beta ) p(1-p) +3 \beta,
\label{l4}
\end{equation}
giving $\lambda_{c,2}=\frac{3}{2} (1 - 3 \beta ) +3 \beta$.

With $\lambda_{c,2}>1 \Longleftrightarrow \frac{1}{2} (1 - 3 \beta )>0  \Longleftrightarrow \beta< \frac{1}{3} \Longrightarrow$ $p_{c,2}$ is a separator, I conclude that $p_{c,1}=0$ and $p_{c,3}=1$ are attractors, which means that in the range $\beta< \frac{1}{3}$ the dynamics stays qualitatively unchanged being similar to the dynamics of the LMM with a slowing down of the dynamics towards the attractors. However, as soon as $\beta>\frac{1}{3}$, qualitative change occurs with $p_{c,2}$ becoming an attractor and $p_{c,1},p_{c,3}$ separators. 

The flow of opinion has been reversed as with the CMM. However here, $p_{c,1}$ and $p_{c,3}$ stay located at $0$ and $1$ contrary to the CMM. The difference lies in the changing direction of the flow of the dynamics when $\beta> \frac{1}{3}$. It is thus interesting to note that giving a probability for one agent to convince two others produces has no effect in the range $\beta< \frac{1}{3}$. The change of $p_{c,2}$ from being a separator to an attractor occurs via $\beta= \frac{1}{3}$, which turns the EMM to a voter model as seen from Eq. (\ref{a3}).

\subsection{Application to Sznajd models}

I apply now the GUF to several versions of Sznajd model, each version  having an update rule built on a different social mechanism.

\subsubsection{The Original Outflow Model (OOM) \cite{sn1}}
\label{oom}

The original Sznajd model considers a one dimensional chain of agents $S_{l,t}=\pm1$, whose opinions are updated according to an update rule inspired from the wisdom principle  ``United we stand divided we fall" \cite{sn1}. The dynamics is implemented by selecting groups of four neighbors. In a given group, the state of the middle pair determines the choices of the two external neighbors. The dynamics is thus outflow and that direction is argued to be socially more realistic that the usual inflow dynamics used for instance with Glauber dynamics where the surrounding spins which influence the central one.

Choosing randomly 4 nearest neighbor agents $S_{1,t}, S_{2,t}, S_{3,t}, S_{4,t}$, the update rule selects the middle pair $S_{2,t}, S_{3,t}$ to update the states of the two external agents $S_{1,t}, S_{4,t}$. The rule  operates as follows:

\begin{enumerate}
\item $S_{2,t} =S_{3,t} \rightarrow S_{1,t+1}= S_{4,t+1}= S_{2,t+1} =S_{3,t+1}=S_{2,t}=S_{3,t}$. 
\item $S_{2,t} =-S_{3,t} \rightarrow S_{1,t+1}=-S_{2,t+1} =-S_{2,t}$ and $S_{4,t+1}=-S_{3,t+1}=-S_{3,t}$ ,
\end{enumerate}
making $r=4$ and $r_u=2$.

A modified version reduces above rules to three agents \cite{sn2}. A pair is chosen to influence one of its two neighbor, either the left or the right one with equal probabilities making $r=3$ and $r_u=1$. Indeed, choosing always either the left or the right agent does modify the results besides doubling the relaxation time to reach equilibrium. Applying the GUF scheme selecting always the right sided agent gives:
\begin{itemize} 
\item $a_1=+ +(+) \rightarrow b_1=+ +(+)$, with $k_1=1$ ,
\\ \item $a_2=+ +(-)  \rightarrow  b_2=+ +(+) $, with $k_2=1$ ,
\\ \item $a_3=+ -(+) \rightarrow  b_3=+ -(+)$, with $k_3=1$ ,
\\ \item $a_4=- +(+)  \rightarrow  b_4=- +(-) $, with $k_4=0$ ,
\\ \item $a_5=- - (+)   \rightarrow  b_5=- - (-)  $, with $k_5=0$ ,
\\ \item $a_6=- +(-) \rightarrow  b_6=- +(-)$, with $k_6=0$ ,
\\ \item $a_7=+ -(-)\rightarrow  b_7=+ -(+)$, with $k_7=1$ ,
\\ \item $a_8=- -(-)  \rightarrow  b_8=- -(-) $, with $k_8=0$ , 
\end{itemize}
which gives $a=0, b=0, c=1, d= 0$. Plugging those values in Eq. (\ref{u3}) gives,
\begin{equation}
p_{t+1}=p_t ,
\label{a4}
\end{equation}
which shows that within the GUF the original Sznajd model is indeed identical to the Voter Model  \cite{voter}. The same finding was already found analytically and numerically by Behera and Schweitzer \cite{frank-voter}.

\subsubsection{The Modified Outflow Model (MOM) \cite{sn2}}

The original Sznajd model generates an antiferromagnetic like ordering when the central pair is ``divided". Now, when $S_{1,t}=-S_{2,t}$, agent $S_{3,t}$ stays unchanged with $S_{3,t+1}=S_{3,t}$. With still $r=3$ and $r_u=1$,  above eight configurations become:
\begin{itemize} 
\item $a_1=+ +(+) \rightarrow b_1=+ +(+)$, with $k_1=1$ ,
\item $a_2=+ +(-)  \rightarrow  b_2=+ +(+) $, with $k_2=1$ ,
\item $a_3=+ -(+) \rightarrow  b_3=+ -(+)$, with $k_3=1$ ,
\item $a_4=- +(+)  \rightarrow  b_4=- +(+) $, with $k_4=1$ ,
\item $a_5=- - (+)   \rightarrow  b_5=- - (-)  $, with $k_5=0$ ,
\item $a_6=- +(-) \rightarrow  b_6=- +(-)$, with $k_6=0$ ,
\item $a_7=+ -(-)\rightarrow  b_7=+ -(-)$, with $k_7=0$ ,
\item $a_8=- -(-)  \rightarrow  b_8=- -(-) $, with $k_8=0$ ,
\end{itemize}
which yield $a=-2, b=3, c=0, d= 0$. Plugging those values in Eqs. (\ref{u5}) gives, 
\begin{equation}
p_{t+1}=-2p_t^3+3p_t^2 ,
\label{a5}
\end{equation}
which is identical to Eq. (\ref{a1}) showing that this modified Sznajd model (MOM) is identical to the Galam Majority Model with $r=3$.

\subsubsection{The Modified Inflow Model (MIM) \cite{sn2, red2}}

In addition to above modified Sznajd model (MOM), another modified version (MIM) has been suggested to account for inflow dynamics (MIM) instead of outflow dynamics \cite{sn2, red2}. In the MIM, the update operates on the central spin $S_{2,t}$ as a function of the states of its two neighbor $S_{1,t}$ and $S_{3,t}$. If $S_{1,t}=S_{3,t}$, $S_{2,t+1}= S_{1,t}=S_{3,t}$. Otherwise $S_{2,t+1}= S_{2,t}$. 
This new rule is motivated by the social principle  ``If you do not know what to do, just do nothing'' principle. This principle is reminiscent of the inertia principle introduced by Galam at a tie in the Local Majority Model with even sizes \cite{voting}. There, the  ``do nothing" is to elect the incumbent candidate in case of a tie vote. 

The associated update rule is obtained again with $r=3$ and $r_u=1$. Above eight configurations become:
\begin{itemize}
 \item $a_1=+ (+) + \rightarrow b_1=+(+)+$, with $k_1=1$ ,
\item $a_2=+ (+) - \rightarrow  b_2=+(+)-$, with $k_2=1$ ,
\item $a_3=+ (-) + \rightarrow  b_3=+(+)+$, with $k_3=1$ ,
\item $a_4=- (+) + \rightarrow  b_4=-(+)+$, with $k_4=1$ ,
\item $a_5=- (-) + \rightarrow  b_5=-(-)+$, with $k_5=0$ ,
\item $a_6=- (+) - \rightarrow  b_6=-(-)-$, with $k_6=0$ ,
\item $a_7=+ (-) - \rightarrow  b_7=+(-)-$, with $k_7=0$ ,
\item $a_8=- (-) - \rightarrow  b_8=-(-)-$, with $k_8=0$ ,
\end{itemize}
yielding $a=-2, b=3, c=0, d= 0$. Those values reproduce the Modified Outflow Model with the same Eq. (\ref{a5}). 

Therefore, within the GUF, both the MOM and MIM are identical and reproduce the LMM.  It should be noted that, while the contributions of the various configurations are different for each model, their addition results in the same expression.  Indeed, the identity between MOM and MIM was already shown in one dimension \cite{in-out}.

\section{A discrepancy}
\label{dis}

At this stage it should be mentioned that  the two rather similar works on the MOM \cite{sn2, red2} derived an exit probability, i.e., the probability for a system with an initial proportion $p_0$ to end up with all agents at $+$,  given by,
\begin{equation}
p_{+}=\frac{p_0^2}{2p_0^2-2p_0+1} ,
\label{a6}
\end{equation}
and all agents at $-$ with probability $(1-p_{+})$. 

Given that in next Section I am claiming that the exit probability is identical to my update rule $p_{t+1}$ without iteration, Eq. (\ref{a6}) exhibits a discrepancy with my Eq. (\ref{a5}) finding. This discrepancy was first discussed in \cite{sn2, pit} and here I am able to solve the issue with the GUF. 

Going back to the implementation of the GUF for MOM,  I could modify the counting arguing that the configurations which do not hold an update are discarded and thus should not be taken into account. The actual process is to pick up a pair of adjacent agents, if they hold different opinions, the pair is discarded. When both agents share the same opinion $+$, if the right sided agent holds the opinion $+$, it keeps it, and if it holds opinion $-$, it shifts to $+$. The same applies for a pair of $-$. 

Therefore I discard configurations $a_3, a_4, a_6, a_7$ keeping only $a_1, a_2, a_5, a_8$ with the respective probabilities,
\begin{itemize}
\item $p_{t,3}=p_{t,4}=p_{t,6}=p_{t,7}=0$ ,  
\item $p_{t,1}= \frac{p^2}{p^2+(1-p)^2} p$ , 
\item $p_{t,2}= \frac{p^2}{p^2+(1-p)^2} (1-p)$ , 
\item $p_{t,5}= \frac{(1-p)^2}{p^2+(1-p)^2} p$ , 
\item $p_{t,8}= \frac{(1-p)^2}{p^2+(1-p)^2} (1-p)$ ,
\end{itemize}
with respectively $k_1=k_2=1$ and $k_5=k_8=0$,and $r_u=1$. The related Eq. (\ref{u1}) writes, 
\begin{equation}
p_{t+1}=\frac{p_t^2}{p_t^2+(1-p_t)^2} ,
\label{a7}
\end{equation}
which is identical to Eq. (\ref{a6}).

It is interesting to notice that above treatment can also apply to the IM. The only difference lies in the configurations contributing to Eq. (\ref{u1}) with discarding $a_2, a_4, a_5, a_7$ and keeping $a_1, a_3, a_6, a_8$. The updated agents is now in the middle of the pair instead of being on the right side. It yields the respective probabilities,
\begin{itemize}
\item $p_{t,2}=p_{t,4}=p_{t,5}=p_{t,7}=0$ ,  
\item $p_{t,1}= \frac{p^2}{p^2+(1-p)^2} p$ , 
\item $p_{t,3}= \frac{p^2}{p^2+(1-p)^2} (1-p)$ , 
\item $p_{t,6}= \frac{(1-p)^2}{p^2+(1-p)^2} p$ , 
\item $p_{t,8}= \frac{(1-p)^2}{p^2+(1-p)^2} (1-p)$ ,
\end{itemize}
with respectively $k_1=k_3=1$ and $k_6=k_8=0$,and $r_u=1$. The corresponding update is identical to Eq. (\ref{a7}).

I thus recover here too the identity of MOM and MIM within the GUF in agreement with \cite{in-out }. However, MOM and MIM are now different from LMM.

\section{Mean field versus the GUF}

Another issue that needs to be tackled is the difference between final states obtained from respectively the GUF and an exit probability as discussed in  \cite{ sn2}. 

The GUF iteration brings the system to an attractor provided enough updates are completed. The result is deterministic. In case, the dynamics is stopped before reaching the attractor due to a limited number $l$ of iterations, the system exhibits a coexistence of a $+$ and $-$ choices in proportions given respectively by $p_{t+l}$ and $(1-p_{t+l})$ if the process started at time $t$. A majority and a minority coexist.

In contrast, the exit probability $p_+$ gives the probability that the system ends up at unanimity along $+$, otherwise unanimity  is along $-$ with probability $(1-p_+)$  \cite{red1, tim}. No intermediate coexisting phase is obtained. It is worth to note that exit probabilities have been derived through elaborated and long calculations \cite{ sn2}. In addition, the probabilistic outcomes have been confirmed with
Monte Carlo simulations \cite{sn2}. The possibility that indeed the system did not reach equilibrium has been evoked  \cite{pit}.

At this stage, I propose an explanation to reconcile the GUF with the exit probability, relying on the nature of the GUF.

It happens that most researchers perceive the GUF as a mean-filed treatment due to the random distribution of agents in the local groups combined with the repeated reshuffling between updates. Indeed, the fact that every agent can interact in principle with any other one, evokes the mean-field hypothesis that every agent interacts with all the others. 

In other words, a mean-field treatment is a one object approach within an averaged environment. Only the choice of one single agent is investigated including its degrees of freedom, while all the other agents are assumed to have chosen the same choice, the averaged choice of the chosen agent. 

With the two choices $+$ and $-$, a mean-field treatment of the dynamics yields a function, which gives the probability for the chosen agent to be $+$ or $-$ with the entire system choice being identical to its current choice. 

Based on the above definition I can assess that a one update GUF is precisely a mean-field treatment. Accordingly, $p_{t+l}$ is the probability to have the entire system at the attractor $p_{c,3}$ and at $p_{c,1}$ with probability $(1-p_{t+l})$. In case of one single the system reaches it with certainty. That makes $p_{t+l}=p_+$.

Therefore, implementing reshuffling with additional updates goes beyond the single site mean-field treatment. As soon as a second update is performed,  $p_{t+2}$ is no longer the exit probability and becomes the proportion of $+$ with $1-p_{t+2}$ being the proportion of $-$. By so doing, for each update the GUF accounts for local fluctuations and the following reshuffling erases the short range correlations which have resulted from the local update. The procedure is in the spirit of real space renormalization group technics. 

It is worth stressing that applying repeated reshuffling to the regular two dimensional Ising ferromagnetic nearest neighbor interactions has exhibited a clear different behavior from the corresponding mean field treatment \cite{reshu-i, reshu}. It is also different from the exact treatment without reshuffling. 

\section{Conclusion}

Within the GUF I have shown that inflow dynamics using the Sznajd model does not impact the outflow dynamics, both the MOM and the MIM are indeed identical and reproduce the LMM with iterations but not within mean-field. In addition I found that the OOM is identical to a voter model as demonstrated in \cite{frank-voter}.  

Those findings are in contradiction with Stauffer's quotation: ``However, the Sznajd model takes into account the well-known psychological and political fact that ``United we stand, divided we fall"; only groups of people having the same opinion, not divided groups, can influence their neighbours. In contrast to the other consensus models, the Sznajd model as published thus far deals only with communication between neighbours, not between everybody. It is a ``word-of-mouth" model."  \cite{stauffer-convince}.

According, putting forward the different social principles:
\begin{itemize}
 \item United we stand divided we fall (MOM),
 \item If you do not know what to do, just do nothing' (MIM), 
 \item Follow the opinion of anybody else (VM),
 \item Follow the majority (LMM),
\end{itemize}
to validate a specific local update can be misleading.  Indeed,``United we stand divided we fall" (MOM) and``Follow the opinion of anybody else" (VM) yield the same dynamics as ``If you do not know what to do, just do nothing'' (MIM) and  ``Follow the majority" (LMM). 

To conclude, the Global Unifying Frame was shown to create a universal tool to investigate any two-state local dynamics, which in turn allows a comparison between models and their related social features. In particular, the GUF provides a key to avoid wrong  claims about the validity of specific psycho-sociological principles. 

Last, but not least, I have developed the GUF for two-state opinion dynamics models but it could be generalized to $3$-state opinions and more, although that will be a tedious task to complete.


\begin{thebibliography}{99.}

\bibitem{brazil}  R. Brazil, The physics of public opinion, Physics World, January issue (2020)

\bibitem{frank} F. Schweitzer, Sociophysics, Physics Today {\bf 71}, 40-47 (2018) 

\bibitem{book} S. Galam, Sociophysics: A Physicist's Modeling of Psycho-political Phenomena, 
Springer (2012)

\bibitem{fortu} C. Castellano, S. Fortunato and V. Loreto, Statistical physics of social dynamics, {\it Rev. Mod. Phys.} {\bf 81}, 591--646 (2009) 

\bibitem{bikas} B. K. Chakrabarti, A. Chakraborti and A. Chatterjee (Eds.), Econophysics and Sociophysics: Trends and Perspectives, Wiley-VCH Verlag (2006)

\bibitem{rum} J. Rabajante and R. E. Umali, A mathematical model of rumor propagation for disaster management J. Nature Studies \textbf{10}, 61-70 (2011)

\bibitem {stauffer-sn} D. Stauffer, A. O. Sousa, and S. Moss de Oliveira, Generalization to square lattice of Sznajd sociophysics model, Inter. J. Mod. Phys. C \textbf{11}, 1239-1245 (2000)

\bibitem {uni} S. Galam, Local dynamics vs. social mechanisms: A unifying frame, Europhys. Lett. \textbf{70}, 705 (2005)

\bibitem{c1}  A. Flache, M. M\"as, T. Feliciani, E. Chattoe-Brown, G. Deffuant, S. Huet, and J. Lorenz, Models of social influence: Towards the next frontiers. J. of Artificial Societies and Social Simulation, \textbf{20(4)}, 2 (2017)

\bibitem{c2} Kozitsin, I.V. A general framework to link theory and empirics in opinion formation models. Sci Rep \textbf{12}, 5543 (2022) 

\bibitem{c3} Devia, C.A., Giordano, G. A framework to analyze opinion formation models. Sci Rep \textbf{12}, 13441 (2022)

\bibitem {voting} S. Galam,  Majority rule, hierarchical structures, and democratic totalitarianism: A statistical approach, J. of Math. Psychology \textbf{30}, 426  (1986)

\bibitem {chopard} S. Galam, B. Chopard, A. Masselot and M. Droz, Competing species dynamics: Qualitative advantage versus geography, Eur. Phys. J. B \textbf{4}, 529-531 (1998) 

\bibitem{mino} S. Galam,  Minority Opinion Spreading in Random Geometry,  Eur. Phys. J.  {\bf B 25} Rapid Note 403-406 (2002)

\bibitem{contra} S. Galam, Contrarian Deterministic Effects on Opinion Dynamics: The Hung Elections Scenario, Physica A 333, 453-460 (2004)

\bibitem{red1} M. Mobilia, S. Redner, Majority versus minority dynamics: Phase transition in an interacting two-state spin system, Phys. Rev. E \textbf{68}, 046106  (2003)

\bibitem{sn1} K. Sznajd-Weron and J. Sznajd, Opinion evolution in closed community, Int. J. Mod. Phys. C \textbf{11}, 1157 (2000)

\bibitem{voter} T. M. Liggett,  \textit{Stochastic Interacting Systems: Contact, Voter, and Exclusion Processes}, Springer, Berlin (1999).

\bibitem{frank-voter} L. Behera and F. Schweitzer, Nonlinear voter models: the transition from invasion to coexistence, Int. J. Mod. Phys. C {\bf 14}, 1331(2003)

\bibitem{sn2} F. Slanina, K. Sznajd-Weron, and P. Przybyla, Some new results on one-dimensional outflow dynamics,
EuroPhys. Lett.,  \textbf{82}, 18006 (2008)

\bibitem {red2} R. Lambiotte and S. Redner, Dynamics of non-conservative voters, Eur. Phys. Letters,  \textbf{82}, 18007 (2008)

\bibitem {in-out} C. Castellano and R. Pastor-Satorras, Irrelevance of information outflow in opinion dynamics models, 
Phys. Rev. E \textbf{83}, 016113 (2011)

\bibitem {pit}  S. Galam and A. C. R. Martins, Pitfalls driven by the sole use of local updates in dynamical
systems, Europhys. Lett, 95, 480 (2011) 

\bibitem {tim}  A. M. Timpanaro and S. Galam, Analytical expression for the exit probability of the q-voter model in one dimension, Phys. Rev. E 92, 012807 (2015)


\bibitem {reshu-i}  S. Galam and A. C. R. Martins, Two-dimensional Ising transition through a technique from two-state opinion-dynamics models,  Phys. Rev. E 91, 012108 (2015)

\bibitem {reshu}  A. O. Sousa, K. Malarz, and S. Galam, Reshuffling spins with short range interactions: When sociophysics produces physical results, Inter. J. Mod. Phys. C \textbf{16},1507  (2005) 

\bibitem{stauffer-convince} D. Stauffer: How to Convince Others? Monte Carlo Simulations of the Sznajd Model, In The Monte Carlo Method in the Physical Sciences, vol 690,  AIP Conference Proceedings 147-155  (2003)





\end{thebibliography}
\end{document}